\documentclass[aps,prc,preprint,superscriptaddress]{revtex4-1}

\usepackage{bm}
\usepackage{amsmath}
\usepackage{epstopdf}
\usepackage{graphicx}
\usepackage{float}
\usepackage{color}
\usepackage{gensymb}

\begin{document}


\title{Multinucleon transfer with time-dependent covariant density functional theory}

\author{D. D. Zhang}
\affiliation{State Key Laboratory of Nuclear Physics and Technology, School of Physics, Peking University, Beijing 100871, China}

\author{D. Vretenar}
\email{vretenar@phy.hr}
\affiliation{Physics Department, Faculty of Science, University of Zagreb, 10000 Zagreb, Croatia}
\affiliation{State Key Laboratory of Nuclear Physics and Technology, School of Physics, Peking University, Beijing 100871, China}

\author{T. Nik\v si\' c}
\affiliation{Physics Department, Faculty of Science, University of Zagreb, 10000 Zagreb, Croatia}
\affiliation{State Key Laboratory of Nuclear Physics and Technology, School of Physics, Peking University, Beijing 100871, China}

\author{P. W. Zhao}
\email{pwzhao@pku.edu.cn}
\affiliation{State Key Laboratory of Nuclear Physics and Technology, School of Physics, Peking University, Beijing 100871, China}

\author{J. Meng}
\email{mengj@pku.edu.cn}
\affiliation{State Key Laboratory of Nuclear Physics and Technology, School of Physics, Peking University, Beijing 100871, China}


\date{\today}

\begin{abstract}
The microscopic framework of time-dependent covariant density functional theory is applied to study multinucleon transfer reactions, with transfer probabilities calculated using the particle number projection method. It is found that similar total cross sections are obtained with two different relativistic density functionals, PC-PK1 and DD-ME2, as well as with the Skyrme functional SLy5 in a previous study, for multinucleon transfer in the reactions: $^{40}{\rm Ca}+{}^{124}{\rm Sn}$ at $E_{\rm lab} = 170$ MeV, $^{40}{\rm Ca}+{}^{208}{\rm Pb}$ at $E_{\rm lab} = 249$ MeV, and $^{58}{\rm Ni}+{}^{208}{\rm Pb}$ at $E_{\rm lab} = 328.4$ MeV. We report the first microscopic calculation of total cross sections for the reactions: $^{40}{\rm Ar}+{}^{208}{\rm Pb}$ at $E_{\rm lab} = 256$ MeV and $^{206}{\rm Pb}+{}^{118}{\rm Sn}$ at $E_{\rm lab} = 1200$ MeV. Compared to the results obtained with the GRAZING model, the cross sections predicted by the time-dependent covariant density functional theory are in much better agreement with data, and demonstrate the potential of microscopic models based on relativistic density functionals for the description of reaction dynamics.
\end{abstract}


\maketitle
\section{Introduction}\label{sec1}
The study of multinucleon transfer (MNT) reactions provides a unique and detailed perspective on nuclear structure and dynamics. This knowledge has important implications for various subfields of nuclear physics, 
nuclear astrophysics, nuclear reactions, reaction mechanisms, and the production of exotic isotopes. 
Over the last few decades, remarkable achievements have been reported on the synthesis of superheavy nuclei and exotic nuclides far from stability, including neutron-rich nuclei \cite{ThoennessenRPP2013}.
The production of heavy neutron-rich nuclei has primarily been based on fragmentation reactions, although the cross section decreases rapidly as the neutron number increases \cite{KurtukianPRC2014}.
Fusion reactions have been the dominant method employed for the synthesis of superheavy nuclei \cite{HofmannRMP2000,OganessianPRC2004}. 
When stable projectiles are used, fusion reactions predominantly produce superheavy neutron-deficient isotopes because of the curvature of the stability line. To extend the chart of nuclide, to create heavy and superheavy neutron-rich nuclei and ultimately reach the predicted island of stability, it is necessary to explore alternative methods \cite{AdamianEPJA2020}.

It was pointed out that MNT reactions might present advantages in the synthesis of superheavy nuclei, primarily attributed to shell effects \cite{DassoPRL1994,ZagrebaevJPG2005,ZagrebaevPRC2006,ZagrebaevJPG2007,ZagrebaevPRL2008,ZagrebaevPRC2008,ZagrebaevPRC2011,ZagrebaevPRC2013}. This prediction prompted a renewed interest in exploring reactions that involve the transfer of multiple nucleons (protons and/or neutrons) between two colliding nuclei. Furthermore, it has been suggested that MNT reactions offer a promising approach for producing neutron-rich nuclei with a neutron number of $N=126$ \cite{ZagrebaevPRL2008}.
Subsequent experimental studies have revealed that the magnitude of the cross section for the MNT reaction $^{136}{\rm Xe}$+$^{198}{\rm Pt}$ is significantly larger than that of fragmentation reactions with $^{208}$Pb as target, particularly on the neutron-rich side~\cite{WatanabePRL2015}. This finding supports the feasibility of the MNT approach in synthesizing neutron-rich nuclei in the region close to $N=126$. MNT reactions have thus emerged as a promising avenue for producing heavy neutron-rich nuclei, attracting considerable  experimental interest~\cite{CorradiJPG2009}. In addition, they can be used to investigate how nuclei interact and how nucleon exchange occurs during a collision. These studies provide useful information on nuclear forces and the dynamics of nuclear reactions.

Numerous theoretical models have been developed and applied in studies of multinucleon transfer reactions, including the GRAZING model~\cite{WintherNPA1994,WintherNPA1995,WenPRC2019}, the dinuclear system (DNS) model~\cite{FengPRC2017,NiuPRC2017,ZhuPRC2017,BaoPLB2018,ZhuPRC2018,GuoPRC2019,ZhangYHPRC2023,LiaoPR2023,LiaoPRC2023,ZhaoPRC2023,FengPRC2023}, Langevin equations~\cite{ZagrebaevPRL2008,ZagrebaevPRC2008,ZagrebaevPRC2011,ZagrebaevPRC2013,KarpovPRC2017,SaikoPRC2019}, the improved quantum molecular dynamics (ImQMD) model~\cite{ZhaoPRC2015,LiPRC2016,YaoPRC2017,LiPRC2019,ZhaoPLB2021,ZhaoPRC2022,LiaoPRC2023}, and the time-dependent Hartree-Fock (TDHF) theory~\cite{SimenelPRL2010,SekizawaPRC2013,SekizawaPRC2014,SekizawaPRC2016,SekizawaPRC2017,SekizawaPRC2017-1,WuPRC2019,JiangPRC2020,WuPLB2022}. Nuclear density functional theory (DFT), in particular, successfully reproduces and predicts basic properties for most nuclides on the nuclear chart~\cite{VautherinPRC1972,DechargePRC1980,RingPPNP1996,VretenarPR2005,NiksicPPNP2011,Meng2016}. Its dynamical extension, the time-dependent density functional theory (TD-DFT), presents a general microscopic framework that, without any parameter specifically adjusted to the reaction mechanism, can be employed to explore multinucleon transfer reactions. Using a particular implementation of this framework, the 
TDHF, significant progress has been made in the study of multinucleon transfer reaction dynamics~\cite{SimenelPRL2010,SekizawaPRC2013,SekizawaPRC2014,SekizawaPRC2016,SekizawaPRC2017,SekizawaPRC2017-1,WuPRC2019,JiangPRC2020,WuPLB2022}. Various factors related to the entrance channel properties such as neutron-to-proton ($N/Z$) ratio~\cite{SekizawaPRC2013}, the relative orientation of deformed ions~\cite{SekizawaPRC2013,SekizawaPRC2016,SekizawaPRC2017,WuPRC2019}, and the charge product $Z_PZ_T$~\cite{SekizawaPRC2013}, have been found to affect the outcome of these reactions. Additionally, the inclusion of beyond-mean-field effects in methods such as in the time-dependent random-phase approximation (TDRPA)~\cite{WilliamsPRL2018,GodbeyPRC2020}, and the stochastic mean-field theory (SMF)~\cite{AyikPLB2008,AyikPRC2020,SekizawaPRC2020,AyikPRC2021,AyikPRC2023,AyikPRC2023-1,AyikPRC2023-2}, has greatly enhanced the accuracy with which various observables in MNT reactions can be computed.

The relativistic extension of TD-DFT, the time-dependent covariant density functional theory (TD-CDFT), has recently been developed~\cite{RenPRC2020} and successfully applied to various nuclear processes, including fusion~\cite{RenPRC2020}, alpha cluster scattering~\cite{RenPLB2020}, chirality~\cite{RenPRCL2022}, fission~\cite{RenPRL2022,RenPRC2022,LiPRC2023,LiPRC2023-1,li2023generalized}, and quasi-fission~\cite{zhang2023ternary}, etc. 

In the present work, a new implementation of the TD-CDFT with particle number projection is for the first time applied to 
multinucleon transfer reactions. The paper is organized as follows. In Sec.~\ref{sec2}, an outline of the basic formalism of TD-CDFT is presented, along with a brief description of particle number projection. Numerical details of the self-consistent static and time-dependent calculations are included in Sec.~\ref{sec3}. In Sec.~\ref{sec4} we present and discuss the results of a study of the MNT reactions $^{40}{\rm Ca}+{}^{124}{\rm Sn}$ at $E_{\rm lab} = 170$ MeV, $^{40}{\rm Ca}+{}^{208}{\rm Pb}$ at $E_{\rm lab} = 249$ MeV, $^{58}{\rm Ni}+{}^{208}{\rm Pb}$ at $E_{\rm lab} = 328.4$ MeV, $^{40}{\rm Ar}+{}^{208}{\rm Pb}$ at $E_{\rm lab} = 256$ MeV, and $^{206}{\rm Pb}+{}^{118}{\rm Sn}$ at $E_{\rm lab} = 1200$ MeV. The first three are benchmark cases for which the results, in addition to data, are also compared to previous TDHF calculations based on Skyrme functionals. For the latter two, the present study provides the first microscopic calculation of MNT cross sections. Finally, Sec.~\ref{sec5} concludes the paper with a brief summary. 
\section{Theoretical Framework}\label{sec2}

In time-dependent DFT the nuclear wave function is at all times a Slater determinant of occupied single-particle states. 
The time evolution of the single-particle wave function $\psi_k(\bm{r},t)$ is governed by the Dirac equation~\cite{RungePRL1984,vanPRL1999,RenPRC2020}
\begin{equation}
	i\hbar\frac{\partial}{\partial t}\psi_k(\bm{r},t)=\hat{h}(\bm{r},t)\psi_k(\bm{r},t).
\end{equation}
Here, the single-particle Hamiltonian $\hat{h}(\bm{r},t)$ can be expressed as
\begin{equation}\label{eq2}
	\hat{h}(\bm{r},t)=\bm{\alpha}\cdot(\hat{\bm{p}}-\bm{V})+V^0+\beta(m+S),
\end{equation}
where $\bm{\alpha}, \beta$ are the Dirac matrices, and $m$ is the nucleon mass. The scalar potential $S$ and vector potential $\bm{V}$ are determined self-consistently by the time-dependent densities and currents as follows:
\begin{subequations}
	\begin{eqnarray}
		\rho_S(\bm{r},t)&=&\sum_k \bar{\psi}_k\psi_k,\\
		j^{\mu}(\bm{r},t)&=&\sum_k \bar{\psi}_k\gamma^{\mu}\psi_k,\\
		j_{TV}^{\mu}(\bm{r},t)&=&\sum_k \bar{\psi}_k\gamma^{\mu}\tau_3\psi_k.
	\end{eqnarray}
\end{subequations}
For further details, we refer the reader to Refs.~\cite{RenPRC2020,RenPLB2020}. 

In collisions, the total wave function $\Psi(\bm{r},t)$ is a single Slater determinant composed of single-particle wave function, 
\begin{eqnarray}
	\Psi(\bm{r},t)=\frac{1}{\sqrt{A!}}\det\{\psi_k(\bm{r},t)\},
\end{eqnarray}
where $A$ is the total number of nucleons. For a given set of initial conditions, the TD-CDFT describes the most probable path of the collision dynamics. Here we consider the time at which the two fragments, a projectile-like fragment (PLF) and a target-like fragment (TLF), produced in the collision, are completely separated. The wave functions of the PLF and TLF are generally not eigenstates of the particle number operator, but a superposition of states with different nucleon  numbers. To calculate the cross section for the reaction products in each channel, particle number projection is employed~\cite{SimenelPRL2010,SekizawaPRC2013}.

The space is divided into the region $V$, which contains the fragment we are interested in, and the complementary region $\bar{V}$. The particle number projection operator for neutrons $(q=n)$ or protons $(q=p)$ in $V$ reads \cite{SimenelPRL2010,SekizawaPRC2013}
\begin{equation}
	\hat{P}_m^{(q)}=\frac{1}{2\pi}\int_0^{2\pi}d\theta e^{i(m-\hat{N}_V^{(q)})\theta}.
\end{equation}
$\hat{N}_V^{(q)}$ is the particle number operator in the region $V$, defines as
\begin{equation}
	\hat{N}_V^{(q)}=\int_Vd\bm{r}\sum_{i=1}^{N^{(q)}}\delta(\bm{r}-\bm{r}_i)=\sum_{i=1}^{N^{(q)}}\Theta_V(\bm{r}_i),
\end{equation}
where the Heaviside function divides the space 
\begin{eqnarray}
	\Theta_V(\bm{r})=\begin{cases}
		1 &\text{if} \quad\bm{r}\in V,\\
		0 &\text{if} \quad\bm{r}\notin V.
	\end{cases}
\end{eqnarray}
By applying the particle number projection operator to the total wave function $\Psi(\bm{r})$, the specific component with $N$ neutrons and $Z$ protons can be extracted,
\begin{eqnarray}
	|\Psi_{N,Z}\rangle=\hat{P}_N^{(n)}\hat{P}_Z^{(p)}|\Psi\rangle.
\end{eqnarray}
Correspondingly, the probability of the occurrence of a reaction product composed of $N$ neutrons and $Z$ protons, denoted as $P_{N,Z}$, is calculated as
\begin{eqnarray}
	P_{N,Z}=\langle\Psi_{N,Z} |\Psi_{N,Z}\rangle=P_N^{(n)}P_Z^{(p)}.
\label{factor}
\end{eqnarray}
Here, $P_N^{(n)}$ and $P_Z^{(p)}$ are the individual probabilities for $N$ neutrons and $Z$ protons, respectively
\begin{eqnarray}
	P_m^{(q)}=\frac{1}{2\pi}\int_0^{2\pi}e^{im\theta}\det\mathcal{B}^{(q)}(\theta)d\theta,
\end{eqnarray}
where
\begin{eqnarray}
	(\mathcal{B}^{(q)}(\theta))_{ij}=\langle\psi_i^{(q)}|\psi_j^{(q)}(\theta)\rangle,
\end{eqnarray}
and
\begin{eqnarray}
	\psi_j^{(q)}(\bm{r},\theta)=[\Theta_{\bar{V}}(\bm{r})+e^{-i\theta}\Theta_V(\bm{r})]\psi_j^{(q)}(\bm{r}).
\end{eqnarray}

Given specific values for the incident energy $E$ and impact parameter $b$, the probability to observe a reaction product with $N$ neutrons and $Z$ protons in $V$ can be determined, denoted as $P_{N,Z}(E,b)$. The cross section for each channel is computed by integrating over the interval of impact parameters,
\begin{eqnarray}
	\sigma_{N,Z}(E)=2\pi\int_{b_{\text{min}}}^{b_{\text{max}}}bP_{N,Z}(E,b)db,
\end{eqnarray}
where $b_{\text{min}}$ is the minimum impact parameter for which a binary reaction occurs, and $b_{\text{max}}$ is the cutoff impact paramter.

\section{Numerical Details}\label{sec3}
In this work, we analyze the following reactions: $^{40}{\rm Ca}+{}^{124}{\rm Sn}$ at $E_{\rm lab} = 170$ MeV, $^{40}{\rm Ca}+{}^{208}{\rm Pb}$ at $E_{\rm lab} = 249$ MeV, $^{58}{\rm Ni}+{}^{208}{\rm Pb}$ at $E_{\rm lab} = 328.4$ MeV, $^{40}{\rm Ar}+{}^{208}{\rm Pb}$ at $E_{\rm lab} = 256$ MeV, and $^{206}{\rm Pb}+{}^{118}{\rm Sn}$ at $E_{\rm lab} = 1200$ MeV. The relativistic density functionals PC-PK1~\cite{ZhaoPRC2010} and DD-ME2~\cite{LalazissisPRC2005} are employed in the TD-CDFT calculation. Before the collision, the projectile and target are in their ground states obtained from stationary self-consistent mean-field calculations~\cite{RenPRC2017,ZhangPRC2022,ZhangIJMPE2023}, with a grid size of $N_x\times N_y\times N_z = 26\times 26\times 26$. The mesh spacing along each axis is $0.8$ fm. In the dynamical case, the grid size is extended to $N_x\times N_y\times N_z = 60\times 26\times 60$. The collision takes place in the $x$-$z$ plane. For the reaction $^{206}{\rm Pb}+{}^{118}{\rm Sn}$, we have extended the space volume by setting the mesh spacing along each axis to $1.0$ fm. For the time evolution of single-particle wave functions, a predictor-corrector method is employed, with a fourth-order Taylor expansion of the time-evolution operator. The time step is $0.2$ fm/$c$. 

At the initial time, the two nuclei are placed on the mesh at a distance $20\sim 24$ fm between them, and it is assumed they initially follow a Rutherford trajectory. After collision, if two distinct fragments are produced, the time-evolution is terminated when the distance between the fragments is larger than $20\sim 24$ fm. If, after the nuclei came into contact, time exceeds $3000$ fm/$c$ without separation, it is considered a fusion event.
For impact parameters larger than $7$ fm, the values $7.5$ fm, $8$ fm, $9$ fm, and $10$ fm are used to compute the cross section. For impact parameters smaller than $7$ fm, the interval of $b$ values is $0.25$ fm. Finally, as the impact parameter approaches $b_{\rm min}$, the interval is narrowed to $0.01$ fm. The cross section is computed with the trapezoidal integration method.

\section{Results and discussion}\label{sec4}
The mean-field equilibrium states of $^{40}{\rm Ca}$, $^{206}{\rm Pb}$, and $^{208}{\rm Pb}$ are spherical when calculated both with the PC-PK1 and DD-ME2 functionals. For $^{58}{\rm Ni}$, both functionals yield a prolate equilibrium shape with the quadrupole deformation parameter $\beta = 0.14$. Without pairing, oblate equilibrium shapes are obtained for $^{118}{\rm Sn}$ and $^{124}{\rm Sn}$, and the corresponding $\beta$ values are 0.04 and 0.06 given by PC-PK1, and 0.05 and 0.11 by DD-ME2. In the case of $^{40}{\rm Ar}$, the equilibrium state is traxially deformed with $\beta = 0.16$ and $\gamma = 47.5\degree$ in the PC-PK1 calculation, while it exhibits an oblate shape with $\beta = 0.16$ for DD-ME2. In collision of deformed nuclei, the initial orientations are not unique. To compare with the results of a previous study in Ref.~\cite{SekizawaPRC2013}, the nuclei $^{124}{\rm Sn}$ and $^{58}{\rm Ni}$ are rotated by $90\degree$ around the $x$ axis, so that their intrinsic deformation symmetry axes are along the $y$-axis for the reactions of $^{40}{\rm Ca}+{}^{124}{\rm Sn}$ and $^{58}{\rm Ni}+{}^{208}{\rm Pb}$. For the other reactions, the deformation symmetry axes are along the $z$ axis.

\begin{figure}[!htbp]
	\centering
	\includegraphics[width=\linewidth]{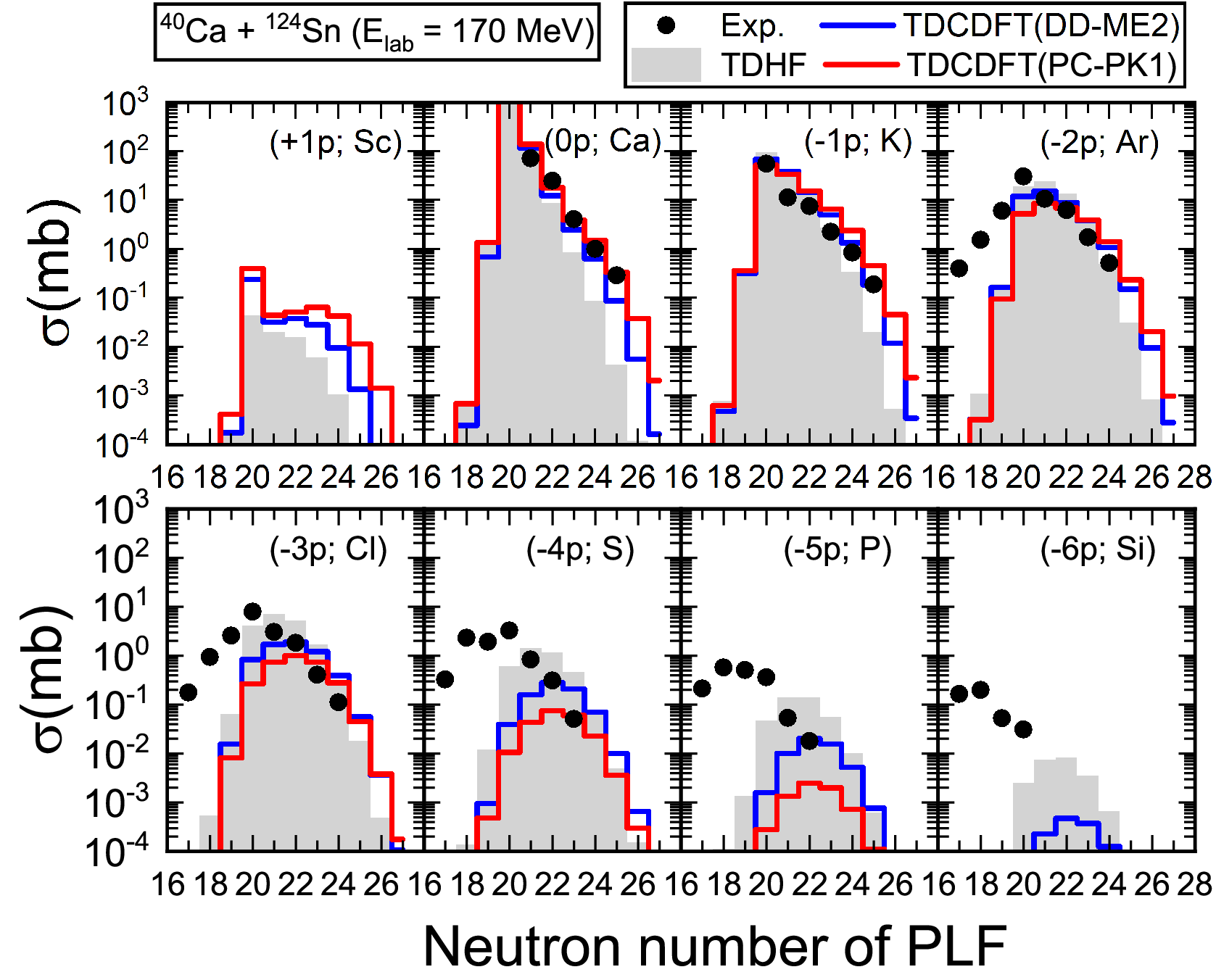}
	\caption{(Color online) Cross sections for the transfer channels to the light fragment in the reaction $^{40}$Ca + $^{124}$Sn at $E_{\text{lab}}$ = 170 MeV. Each panel shows the cross section for a different proton-transfer channel, as a function of the PLF neutron number. Black filled circles denote experimental results from Ref.~\cite{CorradiPRC1996}. Red (blue) histograms represent results obtained by the TD-CDFT calculation with the density functional PC-PK1 (DD-ME2), while grey (shaded) histograms correspond to TDHF results from Ref.~\cite{SekizawaPRC2013}.}
	\label{fig1}
\end{figure}

Figure~\ref{fig1} displays the cross sections for proton transfer channels to the light fragment in the $^{40}$Ca + $^{124}$Sn reaction at $E_{\text{lab}}$ = 170 MeV, from one proton added $(+1p)$ to six protons removed $(-6p)$, as functions of the number of neutrons in the PLF. The terms pick-up and stripping are conventionally referred to the lighter partner of the reaction. The data are from Ref.~\cite{CorradiPRC1996}, and the red (blue) histograms represent results obtained by the TD-CDFT calculation with the density functional PC-PK1 (DD-ME2). For comparison, we also include the cross sections calculated using the TDHF model of Ref.~\cite{SekizawaPRC2013}, with the Skyrme functional SLy5~\cite{ChabanatNPA1998}. 
The largest value of the impact parameter for which fusion takes place is $3.93$ fm in the PC-PK1 calculation, which is close to value $3.95$ fm reported in Ref.~\cite{SekizawaPRC2013}. For the calculation with DD-ME2  this value is $4.24$ fm.

As shown in the figure, the data for the $(0p)$ channel and $(-1p)$ channel are reproduced quite well by all three model calculations. For the $(-2p)$ channel, zero- to four-neutron pick-up channels, in which $^{40}$Ca captures between zero and four neutrons from $^{124}$Sn, are reproduced by theory. The reverse process, however, is underestimated by all three models. Before collision, the neutron-to-proton ratios $N/Z$ for the projectile and target nuclei are $1$ and $1.48$, respectively. One expects that, for collisions between nuclei with different $N/Z$ values, the dominant transfer process is towards charge equilibrium, that is, nucleon transfer tends to equalize the $N/Z$ ratio in the PLF and TLF. Here, this means neutron transfer from $^{124}$Sn to $^{40}$Ca, and proton transfer from Ca to Sn. Nucleons are transferred in the interval of impact parameters close the fusion critical impact parameter. 
The mean values of the primary fragment distribution $P_{N,Z}$, for each impact parameter approach the
equilibrium charge asymmetry values. At low energies, the mean-field approximation provides a good description for the most probable path, with very small dispersion of charge and mass
distributions of primary fragments. In addition, the product form of $P_{N,Z}$ in Eq.(\ref{factor}) indicates that there are no 
correlations between neutron and proton distributions. 
As shown in more detail in Ref.~\cite{SekizawaPRC2013}, the average number of transferred nucleons decreases rapidly as the impact parameter increases. From $(-3p)$ to $(-6p)$, the calculated cross sections decrease faster than the experimental values, especially the ones obtained with the two relativistic density functionals. 
In addition, the peak positions of the theoretical cross sections are shifted to larger average neutron numbers with respect to data. As pointed out in previous studies~\cite{SekizawaPRC2013}, this is most probably due to the fact that the theoretical methods used to calculate cross sections do not take into account neutron evaporation effects. The process of neutron evaporation will naturally shift the cross sections to smaller values of the average neutron number of the PLF, that is, in better agreement with the observed values. We also notice that the cross sections calculated with the three density functionals start to differ significantly only for channels in which a larger number of protons is removed from the projectile. 

\begin{figure}[!htbp]
	\centering
	\includegraphics[width=\linewidth]{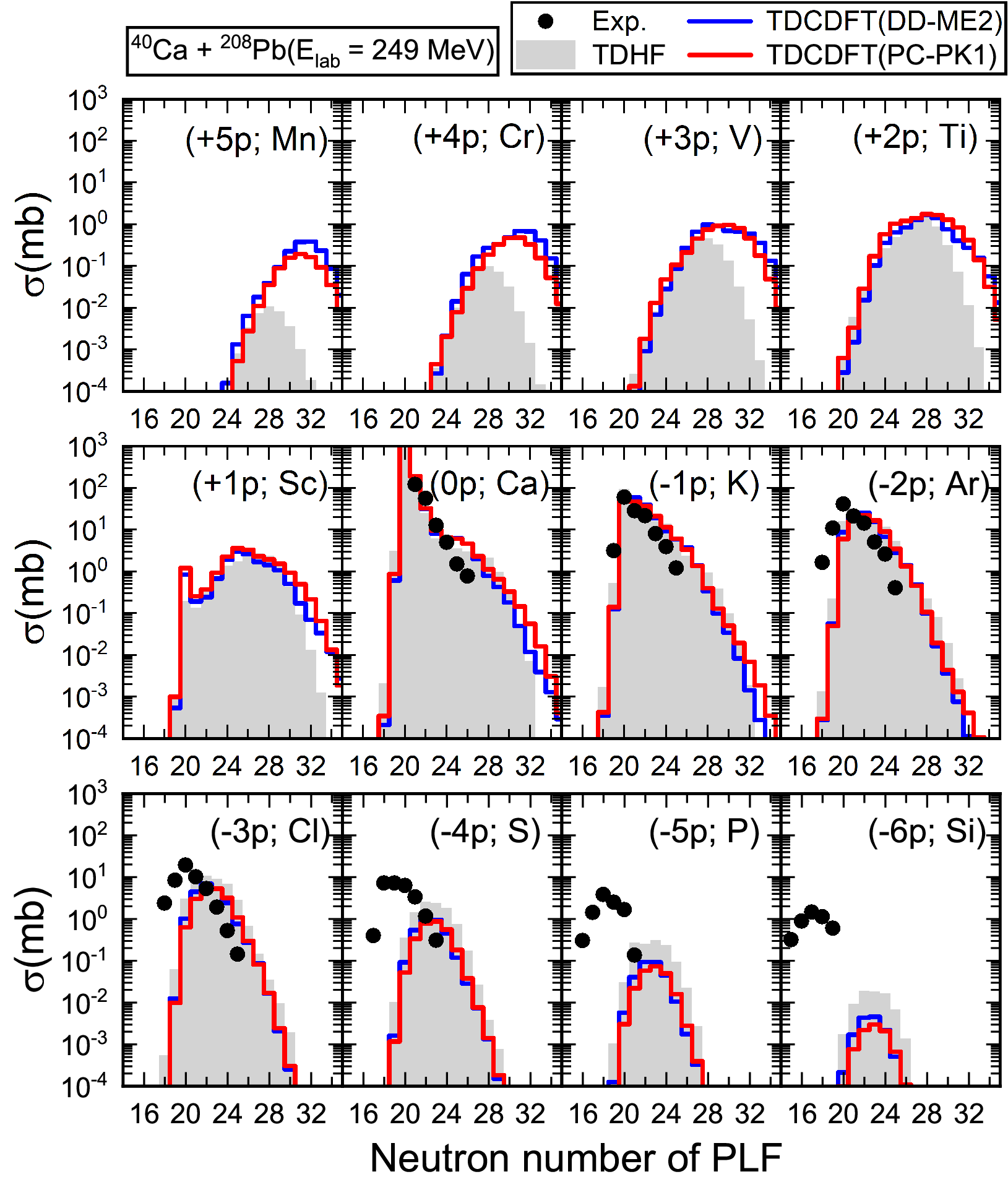}
	\caption{(Color online) Same as in the caption to Fig.~\ref{fig1}, but for the reaction $^{40}$Ca + $^{208}$Pb at $E_{\text{lab}}$ = 249 MeV. The experimental results are from Ref.~\cite{SzilnerPRC2005}.}
	\label{fig2}
\end{figure}

Similar results are obtained for the reaction $^{40}$Ca + $^{208}$Pb at $E_{\text{lab}}$ = 249 MeV. The theoretical cross sections obtained in the present calculation, and in Ref.~\cite{SekizawaPRC2013} with the Skyrme functional SLy5, are compared with available data \cite{SzilnerPRC2005} in Fig.~\ref{fig2}. 
The largest value of the impact parameter for which fusion occurs is $4.63$ fm for the functional PC-PK1, and $4.89$ fm in the calculation with DD-ME2, while the value $4.55$ fm was obtained for Skyrme functional SLy5 in Ref.~\cite{SekizawaPRC2013}. The  three functionals predict comparable cross sections for the proton stripping channels. 
In the $(0p)$ channel, the cross sections for the pick-up of one to four neutrons are well reproduced, while the data for the addition of five and six neutrons in the PLF are slightly overestimated.
Generally, we find that the theoretical results tend to slightly overestimate neutron pick-up and underestimate the neutron stripping cross sections in the $(-1p)$, $(-2p)$, and $(-3p)$ channels. As explained above, this discrepancy can be attributed to the fact that nucleon transfer generally follows the direction of charge equilibrium. In the present  case, the neutron-to-proton ratio of the projectile is $1$, while that of the target is $1.54$, slightly higher than the value $1.48$ for $^{124}$Sn in the previous example. The calculated cross sections for the $(-4p)$, $(-5p)$, and $(-6p)$ channels are considerably reduced, compared to the data. This discrepancy arises from the fusion process occurring before the multi-proton stripping channels are included.

One notices that the cross sections of multi-proton pick-up channels exhibit rather large values, as shown in the panels for $(+1p)$, $(+2p)$, $(+3p)$, $(+4p)$, and $(+5p)$ channels. The primary reason is the occurrence of the quasi-fission process, which leads to the formation of an elongated neck in the binary reaction near the critical impact parameter. When this neck breaks, nucleon transfer takes place in the direction of mass equilibrium, resulting in significant transfer of nucleons to the projectile-like fragment. This is especially pronounced for the two relativistic density functionals used in the present calculation. It is known that fusion reactions occur only with an extra energy push on top of the incident energy when ${\rm Z_P Z_T}$ exceeds a critical value (approximately 1600)~\cite{SimenelEPJA2012}. In the case of the $^{40}$Ca + $^{208}$Pb reaction, the value of ${\rm Z_P Z_T}$ is equal to 1640. Therefore, fusion is hindered in this reaction and instead quasi-fission occurs in the region of small impact parameters.

\begin{figure}[!htbp]
	\centering
	\includegraphics[width=\linewidth]{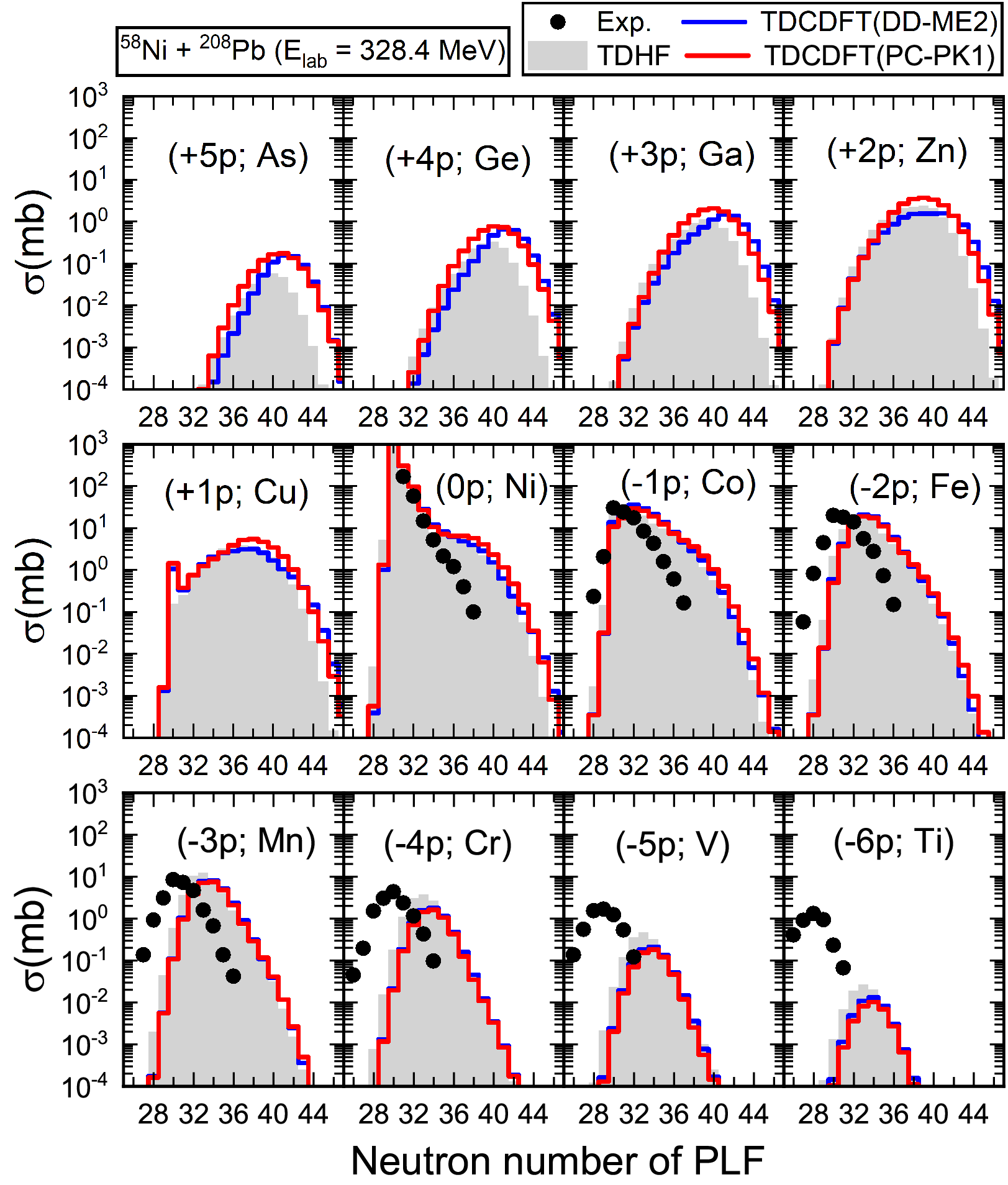}
	\caption{(Color online) Same as in the caption to Fig.~\ref{fig1}, but for the $^{58}$Ni + $^{208}$Pb reaction at $E_{\text{lab}}$ = 324.8 MeV. The experimental cross sections are from Ref.~\cite{CorradiPRC2002}.}
	\label{fig3}
\end{figure}

In Fig.~\ref{fig3}, we plot the theoretical cross sections for the $^{58}$Ni + $^{208}$Pb reaction at $E_{\text{lab}}$ = 324.8 MeV, in comparison with the experimental results of Ref.~\cite{CorradiPRC2002}. In this particular reaction, the value of ${\rm Z_P Z_T}$ is 2296, considerably exceeding the critical value of 1600. 
Consequently, a significant contribution of the quasi-fission process is expected at small impact parameters. The time evolution is traced up to 4000 fm/$c$, to guarantee that the fused system has enough time to eventually separate into two fragments. In the calculation with PC-PK1, the maximum fusion impact parameter for the $^{58}$Ni + $^{208}$Pb reaction is $1.5$ fm, while for DD-ME2 it is $2$ fm. For comparison, a value of $1.38$ fm was determined for Skyrme SLy5 in Ref.~\cite{SekizawaPRC2013}. The critical impact parameter is obviously much smaller than for the previous two reactions, because of the large value of ${\rm Z_P Z_T}$. Quasi-fission takes place at small impact parameters, and thus extends the interval of impact parameters for the calculation of multinucleon transfer cross sections.
In this reaction, the neutron-to-proton ratio of the projectile is $1.07$, while that of the target is $1.54$. Similarly to the previous example, we find that the cross sections for neutron pick-up are slightly overestimated and neutron stripping  underestimated in the $(-1p)$, $(-2p)$, $(-3p)$, and $(-4p)$ channels. For the $(-5p)$ and $(-6p)$ channels, the experimental cross sections are much larger than the calculated ones, and shifted toward smaller neutrons numbers because of neutron evaporation. Multi-nucleon transfer through the neck in the quasi-fission process contributes to the proton pick-up channels, namely the $(+1p)$, $(+2p)$, $(+3p)$, $(+4p)$, and $(+5p)$ channels.

For the above three reactions, the TD-CDFT results for multi-nucleon transfer cross sections obtained with the two relativistic density functionals PC-PK1 and DD-ME2, are consistent with those reported in the TDHF calculation based on the Skyrme functional SLy5~\cite{SekizawaPRC2013}. This comparison validates our model, and in the following we consider the reactions of $^{40}$Ar + $^{208}$Pb and $^{206}$Pb + $^{118}$Sn. The experimental cross sections are from Ref.~\cite{MijatovicPRC2016} and Ref.~\cite{DiklicPRC2023}, respectively, and include data on proton pick-up channels.

\begin{figure}[!htbp]
	\centering
	\includegraphics[width=\linewidth]{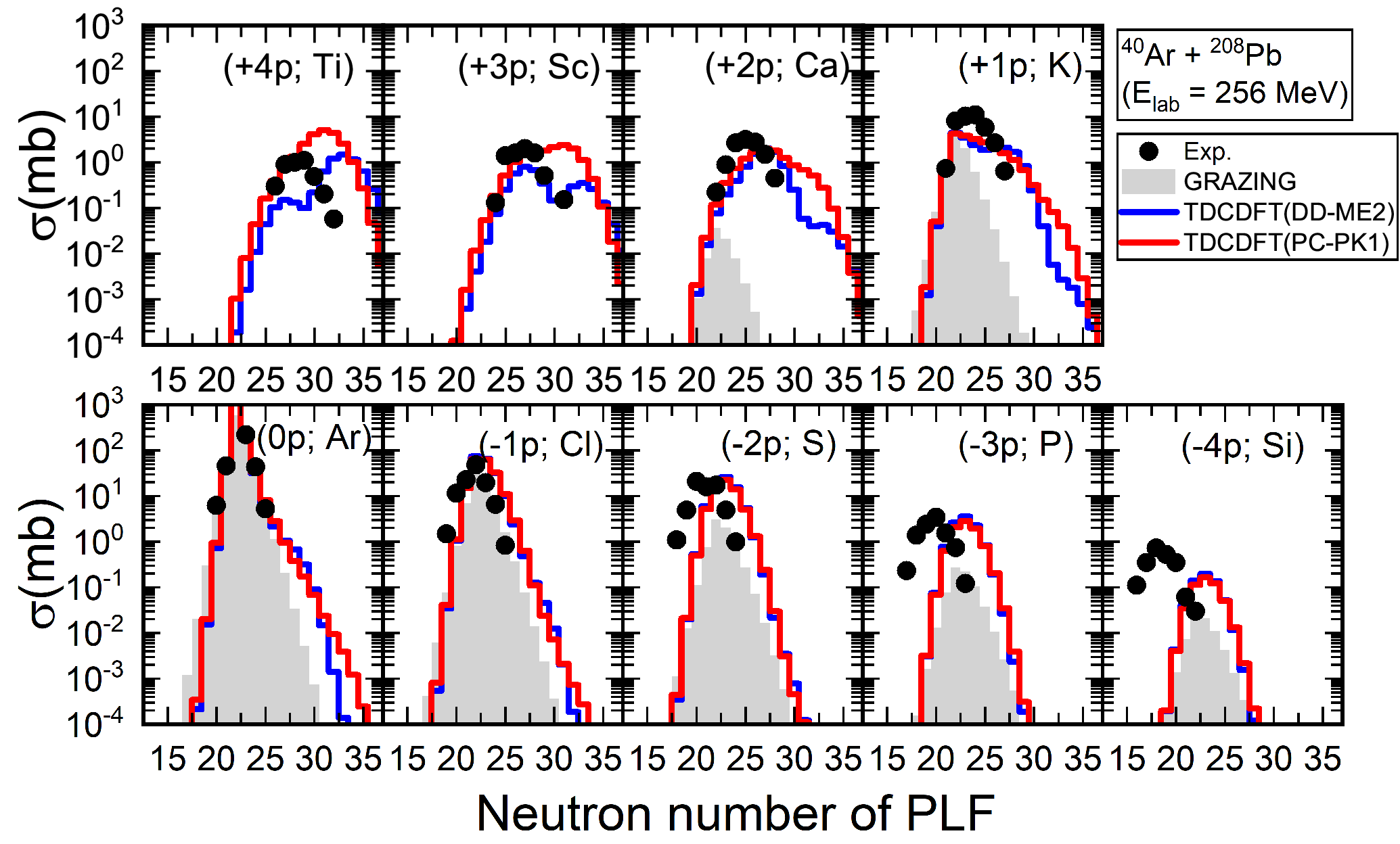}
	\caption{(Color online) Cross sections for the transfer channels to the lighter fragment in the reaction $^{40}$Ar + $^{208}$Pb reaction at $E_{\text{lab}}$ = 256 MeV. Each panel shows the cross section for a different proton- transfer channel, as a function of the PLF neutron number. Black filled circles denote experimental results from 
 Ref.~\cite{MijatovicPRC2016}. Red (blue) histograms denote results obtained in the TD-CDFT calculation with the density functional PC-PK1 (DD-ME2), while the grey (shaded) histograms correspond to GRAZING model calculation~\cite{WintherNPA1994,WintherNPA1995}. }
	\label{fig4}
\end{figure}

Figure~\ref{fig4} displays the cross sections from the $(+4p)$ to the $(-4p)$ channel, for the reaction $^{40}$Ar + $^{208}$Pb at $E_{\text{lab}}$ = 256 MeV. In this reaction, the neutron-to-proton ratio of the projectile is $1.22$, much closer to that of the target ($1.54$) than in the previous three examples. For the functional PC-PK1, the maximum impact parameter for which fusion occurs is $5.99$ fm, while this value is $6.36$ fm for DD-ME2.

In the $(0p)$ channel, the calculation is in excellent agreement with the data, both in the neutron pick-up and neutron stripping channels. For the $(-1p)$ channel, the theoretical results slightly underestimate the experimental cross sections on the neutron stripping side, and overestimate the data for neutron pick-up. As seen in the $(-2p)$, $(-3p)$, and $(-4p)$ channels, this discrepancy gradually becomes larger with increasing the number of protons removed from the projectile, because of neutron evaporation that is not taken into account in model calculations. 
The theoretical results for the $(+1p)$, $(+2p)$, $(+3p)$, and $(+4p)$ channels generally show a very good agreement with the experimental values. In the case of neutron pick-up, however, there is a tendency to overestimate the neutron transfer in the $(+3p)$ and $(+4p)$ channels, which can be attributed to quasi-fission processes involving the transfer of a large number of nucleons.
In our calculation, a rapid increase in the number of transferred nucleons is observed as the impact parameter decreases. Hence, for impact parameters smaller than 7 fm, we reduced the interval of $b$ values from $0.25$ fm to $0.05$ fm. Notably, in the calculation with the functional DD-ME2, the $(+3p)$ and $(+4p)$ channels exhibit a distinct peak associated with quasi-fission, indicating a sudden change in reaction dynamics.
When compared to results of the GRAZING model calculation \cite{WintherNPA1994,WintherNPA1995}, the data for the proton pick-up channels are much better reproduced by the TD-CDFT, especially for the $(+2p)$, $(+3p)$ and $(+4p)$ channels.

\begin{figure}[!htbp]
	\centering
	\includegraphics[width=\linewidth]{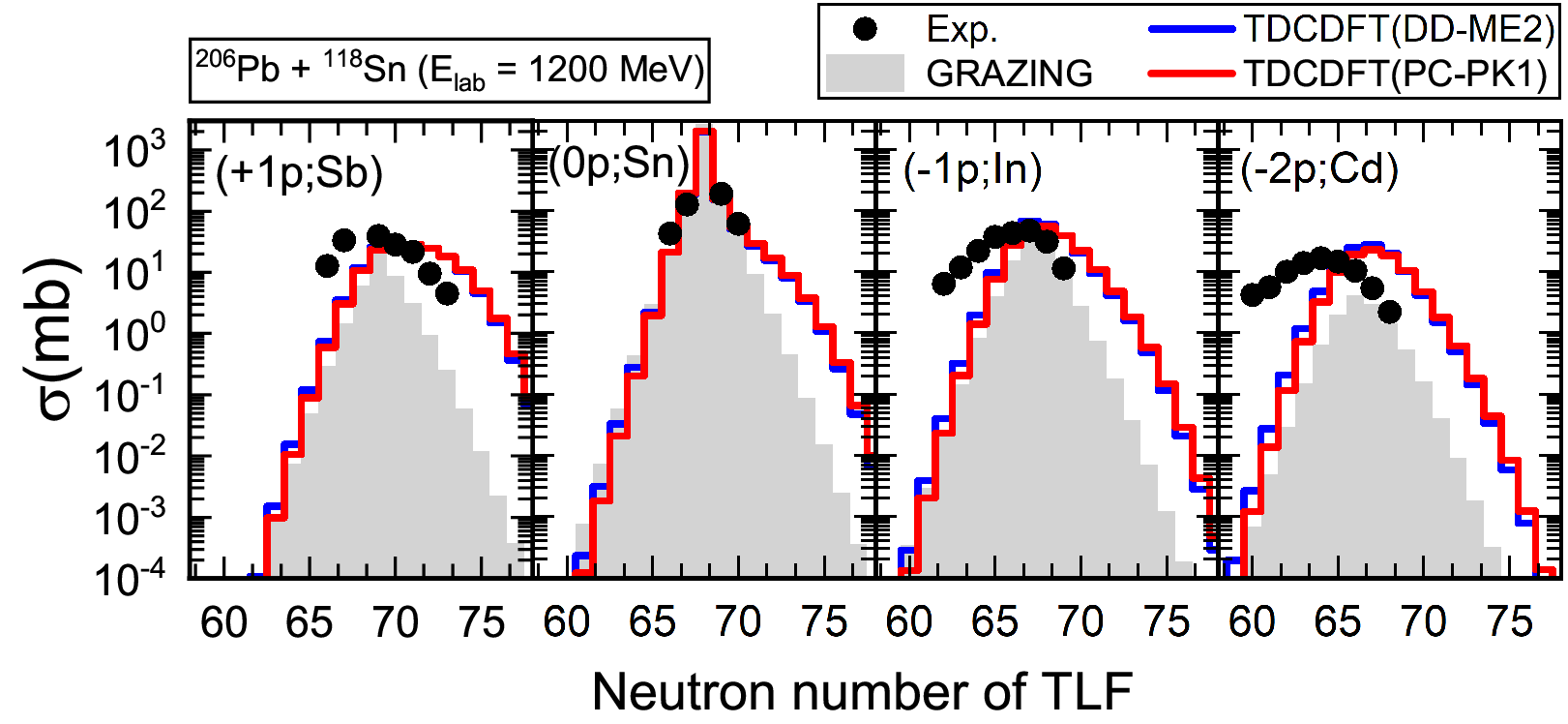}
	\caption{(Color online) Same as in the caption to Fig.~\ref{fig4}, but for $^{206}$Pb + $^{118}$Sn reaction at $E_{\text{lab}}$ = 1200 MeV. The experimental cross sections are from Ref.~\cite{DiklicPRC2023}.}
	\label{fig5}
\end{figure}

The calculated cross sections for the transfer channels from $(+1p)$ to $(-2p)$ to the lighter fragment in the reaction  $^{206}$Pb + $^{118}$Sn at $E_{\text{lab}}$ = 1200 MeV, are compared with the data from Ref.~\cite{DiklicPRC2023}, and to GRAZING model calculation in Fig.~\ref{fig5}. In the $(-2p)$ channel, the magnitude of the cross section is comparable to the data. The position of the peak is at $N=67$, which is three neutrons more than in the experimental data. The neutron-to-proton ratio of the projectile is $1.51$, larger than that of the target at $1.36$. The direction of charge equilibrium, therefore, is the transfer of neutrons from the projectile to the target, while protons are transferred in the opposite direction. The position of the peak in the $(-1p)$ and $(0p)$ channels coincides with the experimental data. For the $(+1p)$ channel, the PC-PK1 and DD-ME2 results are in better agreement with the data compared to the GRAZING model. Even though the peak positions are predicted at larger neutron number, overall the calculated cross sections reproduce the experimental data to a very good level of 
agreement.

\section{Summary}\label{sec5}

The time-dependent CDFT with particle number projection has been applied to a study of multinucleon transfer reactions. By employing two relativistic density functionals PC-PK1 and DD-ME2, we have calculated the MNT total cross sections for five reactions. The results for $^{40}{\rm Ca}+{}^{124}{\rm Sn}$ at $E_{\rm lab} = 170$ MeV, $^{40}{\rm Ca}+{}^{208}{\rm Pb}$ at $E_{\rm lab} = 249$ MeV, and $^{58}{\rm Ni}+{}^{208}{\rm Pb}$ at $E_{\rm lab} = 328.4$ MeV, have been compared to available data, and theoretical cross sections obtained in a previous study based on the Skyrme functional SLy5. These are benchmark examples, here used to validate our model and the particular numerical implementation. 

For all three reactions, in the $(0p)$ and $(-1p)$ channels, the calculated cross sections are found in very good agreement with the data. In the $(-2p)$ and $(-3p)$ channels, the results generally reproduce the magnitude of the experimental cross sections, but tend to underestimate neutron stripping channels. In the $(-4p)$, $(-5p)$ and $(-6p)$ channels, the theoretical cross sections decrease much more rapidly than the data, and the calculated peak positions are shifted to larger neutron numbers with respect to the experimental peaks. It is generally expected that, when neutron evaporation effects are taken into account, the peak positions will shift towards smaller neutron numbers. In general, model calculations reproduce the data for production cross-sections of primary fragments near the equilibrium charge asymmetry values. As primary products are removed further away from the equilibrium charge asymmetry values,  
the rate of decrease of the theoretical cross sections is much faster compared to data. In these channels, the calculated cross sections can be one or two orders of magnitude smaller than their experimental counterparts. An interesting result is that the three density functionals: Skyrme SLy5, the relativistic point-coupling PC-PK1, and the finite-range meson exchange DD-ME2, predict similar MNT cross sections for all three reactions, even though the effective interactions are very different and their parameters were adjusted to nuclear ground-state properties following vastly different protocols. This demonstrates the robustness of the microscopic approach, based on nuclear density functionals, in the description of reaction dynamics. 

In addition, we have also calculated the total MNT cross sections for the reactions:
$^{40}{\rm Ar}+{}^{208}{\rm Pb}$ at $E_{\text{lab}}$ = 256 MeV and $^{206}{\rm Pb}+{}^{118}{\rm Sn}$ at $E_{\text{lab}}$ = 1200 MeV. Recent data on total cross sections have been compared with the results obtained using the TD-CDFT with particle number projection, and the GRAZING model. Generally, the TD-CDFT predicts much better results, in very good  agreement with the experimental cross sections.  

Future studies will include effects of dynamical pairing correlations on MNT, and initial deformations of colliding nuclei. A new implementation of the model will be applied to MNT production of transuranium nuclei, and studies of quasi-fission and fusion-fission reactions with neutron-rich nuclei. 

\begin{acknowledgments}
We thank S. Szilner for useful discussions. This work has been supported in part by the High-end Foreign Experts Plan of China, the National Natural Science Foundation of China (Grants No. 11935003, 11975031, 12070131001, and 12141501), the High-performance Computing Platform of Peking University, the project “Implementation of cutting-edge research and its application as part of the Scientific Center of Excellence for Quantum and Complex Systems, and Representations of Lie Algebras”, PK.1.1.02, European Union, European Regional Development Fund, and by the Croatian Science Foundation under the project Relativistic Nuclear Many-Body Theory in the Multimessenger Observation Era (IP-2022-10-7773). We acknowledge the funding support from the State Key Laboratory of Nuclear Physics and Technology, Peking University. 
\end{acknowledgments}

%

\end{document}